\documentclass[a4paper,twoside]{article}

\usepackage{epsfig}
\usepackage{subfigure}
\usepackage{calc}
\usepackage{amssymb}
\usepackage{amstext}
\usepackage{amsmath}
\usepackage{amsthm}
\usepackage{listings}
\usepackage{multicol}
\usepackage{pslatex}
\usepackage{pgfplots}
\pgfplotsset{width=7cm,compat=1.8}
\usepackage{tikz}
\usetikzlibrary{shapes.geometric, arrows, positioning, calc}
\usepackage{apalike}
\usepackage{SCITEPRESS}     

\tikzstyle{box} = [rectangle, minimum width=4cm, text centered, draw=black]
\tikzstyle{box2} = [rectangle, minimum width=1.5cm, text centered, draw=black]
\tikzstyle{reg} = [rectangle, minimum width=3cm, text centered, draw=black]
\tikzstyle{banner} = [rectangle, text centered, fill=white, draw=none]
\tikzstyle{arrow} = [thick,->,>=stealth]
\begin{document}

\title{Vectorized Character Counting for Faster Pattern Matching}

\author{\authorname{Roman Snytsar}
\affiliation{Microsoft Corp., One Microsoft Way, Redmond WA 98052, USA}
\email{Roman.Snytsar@microsoft.com}
}


\abstract{Many modern sequence alignment tools implement fast string matching using the space efficient data structure called FM-index. The succinct nature of this data structure presents unique challenges for the algorithm designers. In this paper, we explore the opportunities for parallelization of the exact and inexact matches and present an efficient SIMD solution for the $Occ$ portion of the algorithm. Our implementation computes all eight $Occ$ values required for the inexact match algorithm step in a single pass. We showcase the algorithm performance in a multi-core genome aligner and discuss effects of the memory prefetch.}

\onecolumn \maketitle \normalsize \vfill

\section{\uppercase{Introduction}}

FM-index has been developed as a space efficient index for string matching. The backward search over the index finds exact matches of a pattern in time that is linear relative to the length of the pattern regardless the size of the reference. Even though the applications are numerous, for instance text compression \cite{Navarro07}, and indexing \cite{Zhang13}, FM-index has become especially popular with the developers of DNA sequence aligners like Bowtie \cite{Langmead09}, SOAPv2 \cite{Soap09}, and BWA \cite{Li09}. Next, we introduce the fundamentals of FM-index construction and operation.

\section{\uppercase{Background}}

Let $R$ be a string of length $n$ over some alphabet $\Sigma$. A special character $\$$ that is not part of the alphabet and is lexicographically smaller than any character in $\Sigma$ is appended to the end of the string. $R[i]$ denotes a character in R at position $i$, and $R[i, j]$ is a substring of R ranging from $i$ to $j$. Suffix array $SA(R)$ is then defined as an integer array containing starting positions of all suffixes of $R$ in a sorted order so that
\begin{equation}\label{eqSA}
R[SA[i-1], n]<R[SA[i], n], 1<i\le n
\end{equation}
Suffix array could be constructed by simply sorting all suffixes of a string. More sophisticated algorithms take into account the fact that all strings are related to each other and achieve much better asymptotic complexity and practical performance \cite{Puglisi07}.

\begin{figure}[!h]
	\centering
	\setlength{\tabcolsep}{0.4em} 
	\begin{tabular}{lllllllllll}
		\multicolumn{2}{c}{SA}           &  & \multicolumn{2}{c}{BWT}                              &                         & \multicolumn{4}{c}{Occ}                                                                           &                        \\
		&        &  & \multicolumn{1}{l}{}       &                         &                         & A                      & C                      & G                      & T                      &                        \\ \cline{1-1} \cline{5-5} \cline{7-10}
		\multicolumn{1}{|l|}{4} & \$     &  & \multicolumn{1}{r|}{\$ACA} & \multicolumn{1}{l|}{G}  &  & \multicolumn{1}{|l}{0} & \multicolumn{1}{l}{0} & \multicolumn{1}{l}{1} & \multicolumn{1}{l|}{0} &                        \\ 
		\multicolumn{1}{|l|}{0} & ACAG\$ &  & \multicolumn{1}{r|}{ACAG}  & \multicolumn{1}{l|}{\$} &  & \multicolumn{1}{|l}{0} & \multicolumn{1}{l}{0} & \multicolumn{1}{l}{1} & \multicolumn{1}{l|}{0} &                        \\
		\multicolumn{1}{|l|}{2} & AG\$   &  & \multicolumn{1}{r|}{AG\$A} & \multicolumn{1}{l|}{C}  &  & \multicolumn{1}{|l}{0} & \multicolumn{1}{l}{1} & \multicolumn{1}{l}{1} & \multicolumn{1}{l|}{0} &                        \\
		\multicolumn{1}{|l|}{1} & CAG\$  &  & \multicolumn{1}{r|}{CAG\$} & \multicolumn{1}{l|}{A}  &  & \multicolumn{1}{|l}{1} & \multicolumn{1}{l}{1} & \multicolumn{1}{l}{1} & \multicolumn{1}{l|}{0} &                        \\
		\multicolumn{1}{|l|}{3} & G\$    &  & \multicolumn{1}{r|}{G\$AC} & \multicolumn{1}{l|}{A}  &  & \multicolumn{1}{|l}{2} & \multicolumn{1}{l}{1} & \multicolumn{1}{l}{1} & \multicolumn{1}{l|}{0} &                        \\ \cline{1-1} \cline{5-5} \cline{7-10}
		&        &  & \multicolumn{1}{l}{}       &                         &                         &                        &                        &                        &                        &                        \\ \cline{7-11} 
		&        &  & \multicolumn{1}{l}{}       &                         & C= & \multicolumn{1}{|l}{0} & \multicolumn{1}{l}{2} & \multicolumn{1}{l}{3} & \multicolumn{1}{l}{4} & \multicolumn{1}{l|}{4} \\ \cline{7-11} 
	\end{tabular}
	\caption{Data structures comprising FM-index are in boxes.}
	\label{fig:fmindex}
\end{figure}

Burrows-Wheeler Matrix $BWM(R)$ is obtained by writing out all rotations of string $R$ and sorting them lexicographically. The last column of the $BWM$ then forms a string known as the Burrows-Wheeler Transform $BWT(R)$. Sorting string rotations is closely related to sorting prefixes as shown in Figure \ref{fig:fmindex}, and $BWT(R)$ can be easily obtained form $SA(R)$:
\begin{equation}\label{eqBWT}
BWT[i]=\begin{cases}R[SA[i]-1], & SA[i]>0\\
 \$, & SA[i]=0
   \end{cases}
\end{equation}

Next, for each character $b$ in $\Sigma$ and for every $0\le k<n$ we record the number of occurrences of b in the $BWT$ substring $BWT[0, k]$ , and store it in the table $Occ$. Additionally, we store the total of occurrences of all characters lexicographically preceding $b$ in $BWT$ into the table $C$. It is easy to compute $C$ as an exclusive prefix sum of the last row of $Occ$. 
For any single character at position $i$ in a pattern $W$, the interval of rows in the BWM starting with this character is easily computed from $C$:
 \begin{equation}\label{eqInit}
 \{k,l\}=\{C[W[i]], C[W[i]+1]\}
 \end{equation}
 From this initial interval, it is possible to extend the search backward from the starting position using the following recursive procedure:
\begin{small}
	\begin{lstlisting}[mathescape=true,caption={Backward Exact Match.},captionpos=b]
ExactRecur(W, i, k, l)
 if $i<0$ then
  return $\left\lbrace k,l\right\rbrace$
 $b\leftarrow W[i]$
 $k\leftarrow C(b) + Occ(b, k-1) + 1$
 $l\leftarrow C(b) + Occ(b, l)$
 return ExactRecur(W, i-1, k, l)
	\end{lstlisting}
\end{small}

After the search is complete, the final BWT interval is mapped back to the locations in reference using the suffix array.

BWA \cite{Li09} extends this algorithm to allow a predetermined number of mismatches $z$:
\begin{small}
	\begin{lstlisting}[mathescape=true,caption={Inexact Match.},captionpos=b]	
InexRecur(W, i, z, k, l)
 if $z<0$ then
  return $\varnothing$
 if $i<0$ then
  return $\left\lbrace k,l\right\rbrace$
 $I\leftarrow \varnothing$
 $I\leftarrow I \bigcup InexRecur(W, i-1, z-1, k, l)$
 for each $b\in\Sigma$ do
  $k\leftarrow C(b) + Occ(b, k-1) + 1$
  $l\leftarrow C(b) + Occ(b, l)$
  if $k\leq l$ then
   $I\leftarrow I \bigcup InexRecur(W, i, z-1, k, l)$
  if $b = W[i]$ then
   $I\leftarrow I \bigcup InexRecur(W, i-1, z, k, l)$
  else
   $I\leftarrow I \bigcup InexRecur(W, i-1, z-1, k, l)$
 return I
	\end{lstlisting}
\end{small}

Note that every step of the inexact match algorithm consumes eight $Occ$ values compared to two $Occ$ values required by the exact match. In theory, all $Occ$ values could be precomputed, but holding the full $Occ$ array for the human genome reference would consume approximately 100GB of memory. To save memory space, the FM-index over the DNA alphabet is often stored using a cache-friendly approach introduced in \cite{Gog14}, that harkens back to the bucket layout from the original FM-index paper \cite{Ferragina00}. Values of $Occ(*,k)$ for every k that is a multiple of 128 are stored in memory followed by 128 characters of BWT in 2-bit encoding. Four Occ counters occupy 256 bits, as does the BWT string. Such data arrangement aligns well with AVX vector operations. For $Occ$ counters that are not at the factor of 128 positions, the values must be calculated on the fly. 
Furthermore, suffix array is compressed it a similar manner. Only values of $SA[k]$ where $k$ is a multiple of 32 are stored in memory, while all the values in between are recomputed using the Inverse Suffix Array relationships:
\begin{equation}\label{eqISA}
\Psi^{-1}(i)=C[BWT[i]]+Occ(BWT[i], i)
\end{equation}
\begin{equation}\label{eqSAM}
SA[k]=SA[(\Psi^{-1})^j(k)]+j
\end{equation}
It means that the Equation \ref{eqISA} is applied over and over until for some j the result comes out to be a multiple of 32, and the SA value could be constructed according to Equation \ref{eqSAM}.

Even though the memory saving measures do not change the asymptotic complexity of the match algorithm, in reality they add hundreds of computations of $Occ$ to every search. Given that the search is performed multiple times for each and every read out of billions required for the alignment of a human genome, $Occ$ function performance becomes crucial.  

\section{\uppercase{Solution}}
Our approach to computing $Occ$ could be traced back to the algorithm by \cite{Vigna08}, that performs memory table lookups to count character occurrences in each byte of the BWT string. We replace the memory lookups with the half-byte register lookups, building on an idea first proposed by Mula for the bit population count \cite{Mula17}. Note however that we do not attempt to reduce the character counting problem to bit counting, and apply the half-byte technique directly to the BWT string.

The input BWT string is masked with zeros beforehand for situations when the character at position $k$ is in the middle of the byte. The result of 256-bit occurrence count should be corrected for the extra $127-k$ A characters.

\begin{figure}[!h]
	\centering
	\begin{tikzpicture}[node distance=1cm and -1.2cm]
\node (BWT) [reg] {"ccacttgcgaaatttacaaggtttattaggtt"};
\node (A) [reg, below =of BWT] {fa3cfe813f026f45};
\node (B) [reg, below left =of A]	{0a0c0e010f020f05};
\node (C) [reg, below right =of A] {0f030f0803000604};
\node (D) [reg, below =of B] {dfbeaffa7fee7ff7};
\node (E) [reg, below =of C] {8041801141021405};
\node (F) [reg, below =of D] {0dfbeaffa7fee7ff};
\node (G) [reg, below =of E] {0804180114102140};
\node (H) [reg, below =of F] {fdfffefffffeffff};
\node (I) [reg, below =of G] {0000000100000100};
\node (J) [reg, below left =of I] {00000000000007f2};
\node (K) [reg, below =of J] {0000000000000006};

\path (BWT)	edge[arrow] (A)
(A)	edge[arrow] node[banner] {vpand} (B)
edge[arrow] node[banner] {vpsrld, vpand} (C)
(B)	edge[arrow] node[banner] {vpshufb} (D)
(C)	edge[arrow] node[banner] {vpshufb} (E)
(D)	edge[arrow] node[banner] {vpsrlvq} (F)
(E)	edge[arrow] node[banner] {vpsrlvq} (G)
(F)	edge[arrow] node[banner] {vpandnot} (H)
(G)	edge[arrow] node[banner] {vpor} (I)
(H)	edge[arrow] node[banner] {vpsadbw} (J)
(I)	edge[arrow] node[banner] {vpsadbw} (J)
(J)	edge[arrow] node[banner] {vpsubq} (K);
	\end{tikzpicture}
\caption{Example of counting letter 'g' in a 32-character portion of BWT string.}
\label{fig:example}
\end{figure}

\subsection{Lookup}
Every byte in the BWT string is split into its higher and lower half. Since each half byte value cannot be greater than 15, the lookup values now fit into a single vector register and could be retrieved via the VPSHUFB instruction. The lookup returns all four counters in a 2-bit format packed into a byte. Two bits are sufficient as a half byte contains just two characters.
Additionally, the $OccLo$ result is pre-converted into its bit complement to assist with subsequent extraction operation.

\subsection{Extraction}
After the lookup phase, all counters are tightly packed into two vector registers. Any addition operation could result in overflowing the 2-bit values. Before proceeding we have to extract counters for a given character X.

Both $OccLo$ and $OccHi$ are shifted to bring $Occ_X$ into the two lower bits of the byte. High bits of $OccLo$ are then filled with ones and high bits of $OccHi$ – with zeros. At this point $OccHi$ contains unsigned byte values of $OccHi_X$ and $OccLo$ contains values of $255-OccLo_X$.
The merged values are then fed into the VPSADBW instruction. It sums the absolute differences of eight consecutive bytes and stores the result as a 64-bit integer. At the end of this operation the result vector contains four partial sums in the form of 
$2040-\Sigma{(OccLo_X+OccHi_X)}$. The final horizontal sum yields $8160-Occ_X$, and the final subtraction is combinable with the correction for the extra A count.

\subsection{Aggregation}
The extraction sequence runs four times to collect partial sums for characters ACGT, CATG, TGCA, and GTAC in four vector registers. Aggregating four final sums within a single register then takes just three additions and three shuffle operations, only one of which crosses the 128-bit lane boundary. The absence of data dependencies between extraction operations facilitates efficient use of the SIMD pipeline and keeps all arithmetic ports busy.  Outputs for all eight counters required for one step of the inexact search fit into a single AVX512 register and are computed in one vector pass.

\begin{figure}[!h]
	\centering
	\begin{tikzpicture}[node distance=2cm, on grid]
	\node (BWT) [box2] {BWT};
	\node (OccLo) [box2, below left = of BWT, xshift=.6cm] {Lo};
	\node (OccHi) [box2, below right = of BWT, xshift=-.6cm] {Hi};
	\node (Occ12) [box2, below = of OccLo] {G,T,A,C};
	\node (Occ11) [box2, left = of Occ12, xshift=.2cm] {T, G, C, A};
	\node (Occ13) [box2, below = of OccHi] {C,A,T,G};
	\node (Occ14) [box2, right = of Occ13, xshift=-.3cm] {A,C,G,T};
	\node (Occ21) [box2, below right = of Occ11] {G,T,A,C};
	\node (Occ22) [box2, below left = of Occ14] {A, C, G, T};
	\node (Occ31) [box2, below = of $(Occ21)!0.5!(Occ22)$, yshift=.8 cm] {A, C, G, T};
	\node (Lookup) [banner, above = of Occ11, yshift = .8cm] {Lookup};
	\node (Extract) [banner, above = of Occ11, yshift = -.8cm] {Extract};
	\node (Aggregate) [banner, below = of Occ11, yshift = -.6cm] {Aggregate};
	\path
	(BWT)	edge[arrow] (OccLo)
	edge[arrow] (OccHi)
	(OccLo)	edge[arrow] (Occ11)
	edge[arrow] (Occ12)
	edge[arrow] (Occ13)
	edge[arrow] (Occ14)
	(OccHi)	edge[arrow] (Occ11)
	edge[arrow] (Occ12)
	edge[arrow] (Occ13)
	edge[arrow] (Occ14)
	(Occ11) edge[arrow] (Occ21)
	(Occ12) edge[arrow] (Occ21)
	(Occ13) edge[arrow] (Occ22)
	(Occ14) edge[arrow] (Occ22)
	(Occ21) edge[arrow] (Occ31)
	(Occ22) edge[arrow] (Occ31);
	\end{tikzpicture}
	\caption{Data shuffling.}
	\label{fig:aggregation}
\end{figure}
\section{\uppercase{Implementation}}

We have implemented the half-byte $Occ$ algorithm using the AVX2 instruction set. The AVX512 version computing 8 values in parallel has also been integrated in the inexact search algorithm. The assembly code along with the Intel Architecture Code Analyzer throughput report is listed in the Appendix. The code is indeed well balanced across the ports but is expected to bottleneck on the backend meaning that the memory access pattern is crucial for the real word performance. 
\subsection{Experimental Setup}
The computer platform is an Intel Xeon Platinum 8168 system with 16 cores running at 2.7 GHz and 32GB of RAM.
To test the software performance we have run the BWA alignment tool with 16 threads (-t 16) on a 30X Human genome sample NA12878 from the 1000 Genomes database using hg38 as a reference. We have executed the BWA version 7.15 to establish the baseline, and then replaced the $Occ$ code with our AVX2 and AVX512 implementations. The total runtimes got collected from the BWA reports, and the percentage of time spent in the BWA and SA code measured via profiling.

\subsection{Results}

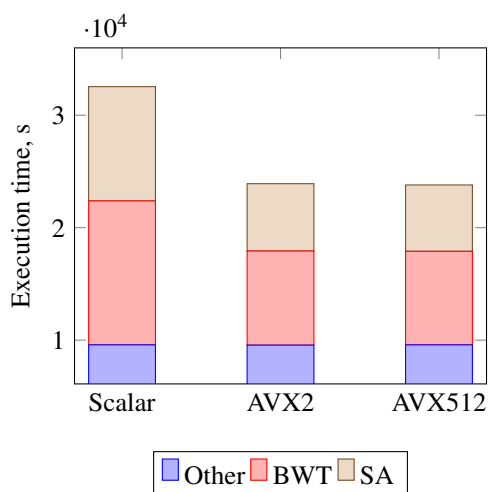
\begin{figure}[!h]
	\centering
		\begin{tikzpicture}
		\begin{axis}[
		ybar stacked,
		bar width=25pt,
		enlargelimits=0.15,
		legend style={at={(0.5,-0.20)},
			anchor=north,legend columns=-1},
		ylabel={Execution time, s},
		symbolic x coords={Scalar, AVX2, AVX512},
		xtick=data,
		]
		\addplot+[ybar] plot coordinates {(Scalar,9580) (AVX2,9555) (AVX512,9579)};
		\addplot+[ybar] plot coordinates {(Scalar,12798) (AVX2,8369) (AVX512,8320)};
		\addplot+[ybar] plot coordinates {(Scalar,10162) (AVX2,5988) (AVX512,5899)};
		\legend{\strut Other, \strut BWT, \strut SA}
		\end{axis}
		\end{tikzpicture}
	\caption{Execution times for NA12878.}
	\label{fig:timing}
\end{figure}

The vectorized SA code runs twice as fast as the scalar version, mostly due to the predictable memory prefetch pattern. The BWT code exhibits only a 30\% speedup, and the switch to AVX512 does not bring any gains. At this point the BWA code is completely memory bound. To utilize the vectorization performance gains completely, we would have to explore ideas for improving the cache locality of the FM-index \cite{Chacon13,Jing13}. Despite the memory bottlenecks, the overall runtime is improved by 25\%.
 
\vfill
\bibliographystyle{apalike}
{\small
\bibliography{SnytsarSimdOcc}}

\section*{\uppercase{Appendix}}

Intel Architecture Code Analyzer report for the AVX implementation of $\Psi^{-1}$.

\onecolumn
\begin{small}
	\begin{verbatim}
Intel(R) Architecture Code Analyzer Version -  v3.0-28-g1ba2cbb build date: 2017-10-23;17:30:24
Analyzed File -  bwa.exe
Binary Format - 64Bit
Architecture  -  SKL
Analysis Type - Throughput

Throughput Analysis Report
--------------------------
Block Throughput: 20.63 Cycles       Throughput Bottleneck: Backend
Loop Count:  22
Port Binding In Cycles Per Iteration:
--------------------------------------------------------------------------------------------------
|  Port  |   0   -  DV   |   1   |   2   -  D    |   3   -  D    |   4   |   5   |   6   |   7   |
--------------------------------------------------------------------------------------------------
| Cycles | 16.3     0.0  | 16.4  |  7.5     7.5  |  7.5     7.5  |  0.0  | 16.3  |  1.0  |  0.0  |
--------------------------------------------------------------------------------------------------

|Num Of|              Ports pressure in cycles            |   |
|  Uops|  0  |  1   |  2  -  D |  3  -  D | 4 |  5  |  6  | 7 |
---------------------------------------------------------------
|   2^ |     | 1.0  | 0.5  0.5 | 0.5  0.5 |   |     |     |   | vpaddq xmm1, xmm10, xmmword ptr [rip+0x21dff0]
|   1  |     |      | 0.5  0.5 | 0.5  0.5 |   |     |     |   | mov r8, qword ptr [rdx]
|   1  |     |      | 0.5  0.5 | 0.5  0.5 |   |     |     |   | vmovdqu xmm2, xmmword ptr [rip+0x21dfd5]
|   1  |     |      | 0.5  0.5 | 0.5  0.5 |   |     |     |   | vmovdqu ymm6, ymmword ptr [rip+0x21dfad]
|   1  |     |      | 0.5  0.5 | 0.5  0.5 |   |     |     |   | vmovdqu ymm7, ymmword ptr [rip+0x21e105]
|   1  |     |      |          |          |   | 1.0 |     |   | vmovq xmm0, r8
|   1  |     |      |          |          |   | 1.0 |     |   | vpbroadcastq xmm0, xmm0
|   1  |     |      |          |          |   | 1.0 |     |   | vpcmpgtq xmm0, xmm0, xmm10
|   2  | 1.0 | 1.0  |          |          |   |     |     |   | vpblendvb xmm3, xmm10, xmm1, xmm0
|   1  | 0.3 | 0.7  |          |          |   |     |     |   | vpandn xmm0, xmm2, xmm3
|   1  | 0.7 | 0.3  |          |          |   |     |     |   | vpsrlq xmm1, xmm0, 0x1
|   1  | 0.3 | 0.7  |          |          |   |     |     |   | vpand xmm9, xmm3, xmm2
|   1  | 1.0 |      |          |          |   |     |     |   | vmovq rcx, xmm1
|   2^ |     |      | 0.5  0.5 | 0.5  0.5 |   |     | 1.0 |   | add rcx, qword ptr [rdx+0x40]
|   1  |     |      |          |          |   | 1.0 |     |   | vpbroadcastb ymm4, xmm9
|   1  |     | 1.0  |          |          |   |     |     |   | vpandn ymm0, ymm4, ymm6
|   1  | 0.7 | 0.3  |          |          |   |     |     |   | vpsrld ymm1, ymm0, 0x17
|   1  |     |      | 0.5  0.5 | 0.5  0.5 |   |     |     |   | vmovdqu ymm5, ymmword ptr [rcx+0x20]
|   1  | 0.3 | 0.7  |          |          |   |     |     |   | vpsrlvd ymm2, ymm5, ymm1
|   1  | 0.7 | 0.3  |          |          |   |     |     |   | vpsrld ymm3, ymm4, 0x1c
|   1  | 0.3 | 0.7  |          |          |   |     |     |   | vpsrad ymm0, ymm7, 0x18
|   1  |     |      |          |          |   | 1.0 |     |   | vpandn ymm1, ymm0, ymm2
|   1  |     |      | 0.5  0.5 | 0.5  0.5 |   |     |     |   | vmovdqu xmm0, xmmword ptr [rip+0x21e01c]
|   1  |     |      |          |          |   | 1.0 |     |   | vpermd ymm2, ymm3, ymm1
|   1  | 0.7 | 0.3  |          |          |   |     |     |   | vpslld ymm8, ymm2, 0x1
|   1  |     |      | 0.5  0.5 | 0.5  0.5 |   |     |     |   | vmovdqu ymm2, ymmword ptr [rip+0x21e01a]
|   1  | 0.3 | 0.7  |          |          |   |     |     |   | vpslld xmm1, xmm4, 0x1
|   1  | 0.7 | 0.3  |          |          |   |     |     |   | vpsubusb xmm1, xmm0, xmm1
|   1  |     |      |          |          |   | 1.0 |     |   | vpmovsxbd ymm3, xmm1
|   1  | 0.3 | 0.7  |          |          |   |     |     |   | vpsllvd ymm0, ymm2, ymm3
|   1  | 0.4 | 0.3  |          |          |   | 0.3 |     |   | vpand ymm4, ymm0, ymm5
|   1  |     |      | 0.5  0.5 | 0.5  0.5 |   |     |     |   | vmovdqu ymm0, ymmword ptr [rip+0x21df5b]
|   1  | 0.3 | 0.4  |          |          |   | 0.3 |     |   | vpand ymm1, ymm4, ymm6
|   1  |     |      |          |          |   | 1.0 |     |   | vpshufb ymm1, ymm0, ymm1
|   1  | 0.7 | 0.3  |          |          |   |     |     |   | vpsrlvd ymm2, ymm1, ymm8
|   1  |     |      | 0.5  0.5 | 0.5  0.5 |   |     |     |   | vmovdqu ymm1, ymmword ptr [rip+0x21df85]
|   1  | 0.3 | 0.7  |          |          |   |     |     |   | vpor ymm5, ymm2, ymm7
|   1  | 0.7 | 0.3  |          |          |   |     |     |   | vpsrld ymm0, ymm4, 0x4
|   1  |     | 0.3  |          |          |   | 0.7 |     |   | vpand ymm3, ymm0, ymm6
|   1  |     |      |          |          |   | 1.0 |     |   | vpshufb ymm2, ymm1, ymm3
|   1  | 0.6 | 0.4  |          |          |   |     |     |   | vpsrlvd ymm0, ymm2, ymm8
|   1  | 0.4 | 0.6  |          |          |   |     |     |   | vpandn ymm3, ymm7, ymm0
|   1  |     |      |          |          |   | 1.0 |     |   | vpsadbw ymm1, ymm3, ymm5
|   2^ | 0.6 | 0.4  | 0.5  0.5 | 0.5  0.5 |   |     |     |   | vpaddd ymm5, ymm8, ymmword ptr [rip+0x21dffe]
|   1  |     |      |          |          |   | 1.0 |     |   | vpermilpd ymm0, ymm1, 0x5
|   1  | 0.4 | 0.6  |          |          |   |     |     |   | vpaddq ymm4, ymm0, ymm1
|   2^ | 0.6 | 0.4  | 0.5  0.5 | 0.5  0.5 |   |     |     |   | vpaddw xmm0, xmm9, xmmword ptr [rip+0x21debc]
|   1  |     |      |          |          |   | 1.0 |     |   | vpmovsxwq ymm1, xmm0
|   2^ | 0.4 | 0.6  | 0.5  0.5 | 0.5  0.5 |   |     |     |   | vpaddq ymm2, ymm1, ymmword ptr [rdx+0x8]
|   2^ | 0.6 | 0.4  | 0.5  0.5 | 0.5  0.5 |   |     |     |   | vpaddq ymm3, ymm2, ymmword ptr [rcx]
|   1  |     |      |          |          |   | 1.0 |     |   | vpermq ymm0, ymm4, 0x4e
|   1  | 0.4 | 0.6  |          |          |   |     |     |   | vpaddq ymm1, ymm0, ymm4
|   1  |     |      |          |          |   | 1.0 |     |   | vmovq xmm0, r8
|   1  | 0.6 | 0.4  |          |          |   |     |     |   | vpsubq ymm2, ymm3, ymm1
|   1  |     |      |          |          |   | 1.0 |     |   | vpermd ymm4, ymm5, ymm2
|   1  | 0.4 | 0.6  |          |          |   |     |     |   | vpcmpeqq xmm1, xmm10, xmm0
|   1  | 0.6 | 0.4  |          |          |   |     |     |   | vpandn xmm0, xmm1, xmm4
|   1  | 1.0 |      |          |          |   |     |     |   | vmovq r8, xmm0
Total Num Of Uops: 65
Analysis Notes:
Backend allocation was stalled due to unavailable allocation resources.
	\end{verbatim}
\end{small}
\vfill
\end{document}